**Super-teams or fair leagues? Parity policies by powerful regulators don't prevent capture**

Adam Sawyer*, Syracuse University, ajsawyer@syr.edu

and Seth Frey, Department of Communication, University of California, Davis, sfrey@ucdavis.edu

*Corresponding author

## Abstract

Much of modern society is founded on orchestrating institutions that produce social goods by fostering motivated teams, pitting them against each other, and distributing the fruits of the arms races that ensue. These "market makers" can be seen as effectively organizing competitive teams into engines of innovation, especially to the extent that techniques and resources diffuse from out of the most innovative teams over time. However, even when the market maker is willing and able to maintain parity between teams, it may fail to maintain a level playing field, as some teams acquire enough advantage within the system to gain influence over it and institutionalize their advantage. Whether this happens can be determined indirectly by investigating if the competitions arranged by a market maker become more deterministic over time. If chance begins to play an equal or greater role in determining contests, that is evidence that the competitive advantages that teams develop are diffusing more quickly than they accumulate. But if the better team tends with increasing consistency to beat worse teams, that is evidence for a divergence in team ability that reflects a potential institutional inequality, as the few top performing teams use their advantage to hoard gains. Using outcomes of over 60,000 games from four leagues and more than 100 years' worth of seasons, we compute the evolving rate of transitivity violations (A > B, B > C, but C > A) to measure the ability of leagues to maintain parity between teams, and support the efficient generation and distribution of innovation. Comparing against a baseline of randomly permuted outcomes, we find that basketball has become less competitive over time, suggesting that teams diverge in performance, and reflecting a possible failure of market makers to tame their overpowered teams. This finding is especially notable given that, in the domain of professional sports, market makers are clearly incentivized to actively maintain a level playing field. Our results suggest that rich-get-richer dynamics are so pernicious that they can even emerge under the watch of a powerful administrator that is motivated to prevent them.

## I. Introduction

Goal-oriented human teams are a major driver of social and economic activity around the world, particularly when they compete. Much of modern society is founded on institutions that produce social goods by pitting motivated teams against each other and distributing the fruits of their unending struggles. These "market makers" — funding agencies, managers of work or sports teams, actual markets — can be seen as effectively orchestrating competitive teams into "innovation engines," as teams' incentives move them to continually innovate, and their advances diffuse through the system.

However, the institutions that emerge to orchestrate team-level competitions do not just play a static puppet master role, they can also evolve in how they structure inter-team contests. For example, imagine an institution that has been organizing multi-team contests for many years. How does that institution's development affect its ability to not just foster, but spread innovation, and advance the performance of the entire system? If the institution and its member teams become more professional, whether in response to more experience, the diffusion of innovation, or the opportunity to gain greater resources, should we expect some teams to preferentially capture and retain their advantages ("the rich get richer") or for the community of teams to improve together over time ("a rising tide lifts all boats")? While these are not necessarily mutually exclusive, an understanding of their relative influences, and trends over time, can inform the ability of the market maker to manage its constituents.

In this study, we observe professionalized basketball teams organized into leagues which have competed in the same round-robin fashion for several decades. These teams have had rotating memberships through this duration but have maintained largely stable corporate identities. During this time, the rules and practices of the league have evolved, its resources and level of professionalization have increased dramatically, and society's level of engagement with the league has increased, attracting an ever greater pool of human capital to its teams, as the teams themselves have become ever more highly skilled. What should be the effect of the league's growth and professionalization on how its teams learn from each other?

We can measure this by asking a more specific question: do the outcomes of contests between teams become more, less, or equally competitive over time? As the ecosystem of teams gains more resources, should member teams diverge from each other in ability, with the best getting better? Should they approach each other in ability, perhaps as play approaches human cognitive limits, and this ceiling effect reduces the role of performance in determining outcomes, relative to noise? Or should arms race dynamics nullify the effects of other changes and keep team performance differences unchanged? This last hypothesis, reminiscent of Louis Carroll's Red Queen's race ("*Now, here, you see, it takes all the running you can do, to keep in the same place*") highlights a connection to evolutionary arms races predicted by evolutionary biology's Red Queen hypothesis (Stenseth and Maynard Smith 1984), but in a business rather than biological evolutionary context (Barnett and Hansen 1996).

By distinguishing these theories within a league of teams, we gain insight into how that league may be driven by inter-team dynamics. If basketball becomes more competitive over time, then innovations in the sport are likely to be diffusing quickly through a system, with team-level competitive advantages apparent only transiently, from season to season. Conversely, if contests are becoming less competitive, innovations would seem to be pooling among some competitors and not diffusing throughout the network, despite any of the league's efforts to the contrary.

## II. Background

### A. Innovation and resource distribution across firms

By intent, market makers protect the environments conducive to the "creative-destruction" envisioned by Schumpeter that leads to innovation, diffusion and growth. Here, we stretch more business-focused conceptions of innovation — such as "the implementation of a new or significantly improved product (good or service), or process, a new marketing method, or a new organizational method in business practices, workplace organization or external relations" (OECD 2005) — to include the development of new strategies, techniques, and practices in excelling at a team sport. In both, innovation promotes attention to efficiency both in process and outcome and can feature as a fundamental condition of wider changes to management, organizational design, and the wider culture of society (Barnett 1953; Abrahamson 1991). The benefits of innovation, however, rests on its ability to diffuse (Robertson, 1967). As stated in the GNU Manifesto,

"creativity can be a social contribution, but only insofar as society is free to use the results" (Stallman, 1985). Market makers try to ensure that the distribution of resources does not become so unbalanced so as to inhibit access to the benefits of innovation (Hsieh and Klenow 2009; Kogan et al. 2017).

Absent an effective system to regulate or incentivize the protection of a competitive market, innovating firms can go to tremendous lengths to retain exclusive access to their advances, even when the result is market failure (Barney 1991). Blind emphasizes that the market makers in an innovation system can be "influenced or pressurised by progress in science and technology" in the long run (2012). Market makers must adapt to innovation over time, as firms within them seek to accumulate resources and exclude competitors from accessing innovations. Otherwise, market makers can become vulnerable to regulatory capture, a process of interest in industrial economics, in which a single market actor, usually a monopolist, gains access to the regulatory process and uses it to institutionalize their advantage.

The problem of runaway concentrations of power is general to social science. In organization science it is evident in fat-tailed distributions of firm size (Simon and Bonini 1958). In economics it can be seen not just in the emergence of monopolists (Encaoua and Jacquemin 1980), but in the kind of income inequalities observed by Vilfredo Pareto (1935). In sociology, researchers in the science of science identified comparable disparities in citation patterns (Dion et al. 2018), and even the degree distributions of online friendship networks (Ahn et al. 2007; Ugander et al 2011). Network scientists have identified dozens of mechanisms that can produce skewed "hub and spoke" networks from balanced networks (Clauset et al. 2015). In political science, runaway power dynamics are the basis of theories like the Iron Law of Oligarchy (Michels 1915), which suggests that democracies are inherently vulnerable to the emergence of a ruling class of oligarchs. Although much work argues that these dynamics are built into "social physics,"(Radicchi et al 2008; Barabási and Albert 1999) one might suppose that a high-powered outside administrator with a motivation to ensure balance among agents would be sufficient to prevent the concentration of power among them. Basketball provides precisely such a case, and its success or failure at managing emergent power dynamics provides a test of whether rich-get-richer dynamics are so endemic to human affairs that they emerge even in systems that explicitly try to design them away.

The market makers that coordinate professionalized sporting entertainment, such as basketball administrative offices, commissioners, and owners' committees, have sought to collaborate to structure the "on-court" and "off-court" rules of the league in ways that protect competitive balance. The annual draft, a system for matching junior players to teams, is structured as a rebalancing matching system that assigns the strongest players to the weakest teams. The NBA's salary control measures have been implemented to incentivize teams to stay within a range of comparable franchise expenditures on player salaries (Hausman and Leonard, 1997). While advancements in fitness regimes and the formalization of data analytics as part of scouting and game preparation have naturally diffused throughout professionalized basketball, some technologies required additional coordination. One such example is the NBA's league-wide introduction of SportVU, an optical tracking technology that assists in performance analysis (Westney, 2013). League administrators across the professional sports industry intervene in these instances to raise the floors and ceilings of possible performance for all teams, and to generate uncertainty, even as teams try to become predictable in their dominance (Merritt and Clauset, 2014).

Still, league administrators face a fundamental tension with the network of team franchises that make up a league, and there is anecdotal evidence that the NBA, for example, has long been compromised by capture. As teams seek to maximize dominance in their favor (at least in the long run) by utilizing innovations for competitive advantage, league administrators seek to maintain fan interest in an ideally uncertain outcome. The risk, in this setting, is that a single team might wield so much influence over, for example, fans and profits, as to gain leverage over league-level decision making, as administrator's perceive short-term commercial benefit from tolerating the dominance of one team, at the long-term cost of the parity that undergirds professionalized athletic contests.

Protecting a sport's competitiveness while remaining efficient in outcomes is the regulatory challenge of league administrators during the long-run innovation cycles of a competitive system. In recent years, league administrators in professional basketball have faced three particular challenges in efforts to maintain competitive balance among a system of competing franchises. First, administrators in basketball have struggled with the advent of dominant teams consisting of top-level individual performers who have decided to pool labor in one destination to ensure a greater probability of success. In some cases, these "super-teams" have been observed to achieve predicted levels of dominance leading to discussions for regulating communication between teams (Wojnarowski and Lowe, 2019). Second, at the other end of the spectrum, less-skilled teams face structural incentives to lose a greater amount of games toward the end of a season in order to gain preferential opportunities in future talent acquisition via the annual draft. Despite the introduction and subsequent reforms of a draft lottery to prevent "tanking" a season, lower-skilled teams seem to systematically lose at a higher rate later in a season (Price et al. 2010). Third, league administrators face a degree of short-term benefits from permitting extended runs of a single team's dominance, contextually evident in the growth of global interest in the NBA during the Bulls 1990 dynasty. However, this comes at the risk of losing fan interest outside of the dominant team's viewer market which would cut into league revenue in the long-run.

### B. The structure of professional basketball

National professionalized basketball began in 1946 with the founding of an eleven-team league called the Basketball Association of America (BAA). After three seasons, the BAA became the National Basketball Association (NBA), and after enduring a ten-year competition with the rival American Basketball Association from 1967 to 1976, the NBA has gone on to become the most prominent professionalized men's basketball league in the world today. The Women's National Basketball Association was founded in 1997 with eight teams and continues to play today as the professionalized basketball league with the second-longest existence behind the NBA. Having access to several semi-independent, and at some points competing, leagues enables us to compare several comparable cases, towards a more general claim.

Despite dramatic changes in rules, including the introduction of the three-point line in 1978, and evolutions in labor mobility through free-agency, the basic structure of professionalized basketball has largely remained the same. Teams have played roughly the same amount of games each season, and unlike most other sports of comparable scale, the round-robin schedule structure has existed from the beginning of professionalized basketball and persists to today: professionalized basketball teams have consistently played every other franchise in the league membership since the founding of the BAA. The well-balanced round-robin structure of professional basketball seasons, with each team in a division playing all others the same number of times, was a major factor in our choice of basketball for this analysis over other professional sports. This is not the case in comparably popular professional sports like baseball and American football (but similar to ice hockey). As opposed to the more tourney-like structure of these other sports, two teams will compete independent of relative skill, and high-performing teams will not be oversampled. The round-robin structure therefore allows direct estimation of the performance difference of most pairs of teams (by directly estimating that performance difference), which has the additional side effect of facilitating comparisons across seasons within leagues, and to some extent across leagues.

Helping to manage the scaling problems of a round-robin design, a square increase in number of games for every unit increase in the number of teams, large basketball leagues are divided into conferences, which are divided further into divisions. Teams in each subunit play a series of preplanned round-robins. Once the regular season is over, the best teams are elevated to league-level playoffs, which follow more of a tournament design. We therefore restrict our attention to regular season play. Other features of basketball make it amenable for the method utilized here. As a national, professionalized, and very popular sport, it has a long history of advancement in tactics, ability, scouting, etc., and has witnessed a large number of observations (games). It is a fast-paced and short game, which supports a very large number of games (observations) in a standard season (currently over 1200). And, basketball games never end in a tie result, which means that every game contributes clear signal.

In the maturation of a professional league over time, the competing interests among the memberships of teams can exert pressure on league administrators. As teams seek to sustain their competitive advantages, how do league administrators protect the competitiveness on which the sport depends?

### III. Method

We test three competing hypotheses using competitive results from four leagues of professional basketball.The dataset we introduce consists of over 60,000 observations between the years 1946 and 2019 that include the home team, the visiting team, and the final score for two contemporary major professional basketball associations (NBA and WNBA) as well as two historical professional associations (ABA and BAA). Seasons which were shortened due to lockout (NBA 1998–99 and NBA 2011–12) or due to the disbandment of an association (ABA 1975–1976) were thrown out due to insufficient games between teams.

Our study imposes a few assumptions on the data. We assume that basketball has changed, and that the basketball teams of today would outcompete comparably ranked basketball teams of the past, whether due to increased resources, skill, or some other consequence of growth and professionalization in the sport. From this starting point we ask whether randomness has come to play a greater, lesser, or equal role in determining game outcomes.

[[Figure 1 here]]

### A. Transitivity Violation Test

To measure the competitiveness of team contests we measure the number of *transitivity violations* over each season's games. For each season, we repeatedly selected groups of three teams at random, and then three games in which each was matched with the other two (Figure 1). Within these triples, if Team A won its game against Team B, who then defeated Team C in their game, transitivity is achieved if A beat C in the third game (A>B>C). If Team A lost to team C, we considered this transitivity violated. One advantage of this measure is that it has natural upper and lower bounds. In a sport in which noise plays no role and skill determines outcomes completely (one akin to running), transitivity violations should be zero. And in a game of chance, transitivity violations should be at a maximum. For example, Rock Paper Scissors, a game of chance, should be maximally intransitive: Whether Player A beat Player B, and B beat C, should offer no information as to whether Player A will beat C. The transitivity violation rate is in this sense a general measure of the determinism of contests.

Our argument is that a change in transitivity violations over seasons is evidence about the game-by-game competitiveness of a sport. However, other factors may impact the rate of transitivity violations over time. To address this,, we developed a baseline to control for the expansion or contraction of the league membership, the organization of teams in divisions, and other season-dependent confounding factors which can also change the rate of transitivity violations across seasons. Therefore, we do not report the observed transitivity violations raw. Instead, we subtracted from the observed rate a season-level "baseline" transitivity violation rate that randomized final scores. To create this intransitivity baseline, we shuffled all results within each season to produce a random dataset of permuted competitions and outcomes, and computed an average baseline transitivity rate over 5000 replicates of this procedure. With this baseline, we can control for confounding cross-season effects on transitivity and focus entirely on within-season effects on transitivity, such as our mechanisms of interest. By this procedure, we found a baseline rate of violations of transitive ordering of around 25% for each season, but with a slightly upward slope, due at least to growth, in all leagues, in the number of teams with basketball's increasing popularity over decades. See Figure S1 in the supplementary section for a direct comparison of the baseline and observed intransitivity rates.

Finally, we subtracted this intransitivity baseline from the observed rate, resulting in a differential measure that accounts for the distance between the slope of the observed rate and the constant rate of transitivity violations in the intransitivity baseline.

To compare the results of the transitivity test against season-by-season results, we also measured the standard deviation of winning percentage of teams in each league across the seasons in the dataset. Unlike win totals, which fluctuate based on the number of competitions and teams organized into a league, winning percentage provides a normalized measure of overall success.

### B. Hypotheses

We the authors could not agree on a hypothesis amongst ourselves, setting this up as an exciting test. Under the first hypothesis, which proposes that leagues cannot prevent teams from diverging in skill over time, we would expect higher levels of transitivity violations in 1946 than in 2019. In other words, the transitive relationship holds true with greater frequency, resulting in greater competitive imbalance. If Team B beats Team C, then Team A beats both Team B and Team C with greater frequency today than in the past.

Under the second hypothesis, which proposes that the market maker manages competitive balance more effectively over time, and skill differentials become smaller, we would expect to see lower rates of transitivity violations in 1946 compared to 2019. This hypothesizes that teams compete under greater parity on a game by game basis, thus creating more illogical, intransitive outcomes that result more from idiosyncratic events than differentials in innovation or skill.

Under the third, the "Red Queen" hypothesis, the rate of transitivity violations should be about constant over time as innovations diffuse across the population of teams at the same rate they accumulate, resulting in relatively little change in the predictivity of the competitive ranking of three randomly chosen teams and matches. This hypothesis would be best supported by a failure to falsify a null hypothesis of zero effect of time on transitivity.

## IV. Results

Figure 2 shows the difference between the observed and baseline transitivity violation rates over time by league. We found that, relative to baseline, the difference in the transitivity violation rate has increased. In a random draw of three teams, a dominant team emerges more often. This result was based on a linear regression with the difference in observed and baseline transitivity violation rates as the dependent variable, a covariate for season and dummy variables for league (Table 1). Standard errors are robust for heteroskedasticity. Professionalized basketball has become less competitive over time, as measured by the number of seasons since the league's inauguration ($t(94)$=–2.053; $p<0.05$; Figure 2, Table 1). In addition to the overall decrease in transitivity violations over time, we find that the observed rate of transitivity violations was lower than the baseline for all leagues. This finding is consistent with the (already undisputed) belief that basketball is more than a game of chance, serving as a sanity or face validity check for the overall suitability of our problem formulation.

To explore alternative explanations of our result, and establish its robustness, we specified a second model (Table 1, second column) that replaced a league's season number with another time variable: the calendar year when the season was played. In the second model this other temporal covariate also correlates negatively with transitivity violation rate. However, in a post hoc third model combining both calendar year and league age, the two temporal variables seem to become redundant and the WNBA variable drops from the model due to a negative multicollinear relationship with the NBA (correlation = –0.753). If there were enough diffusion from the oldest leagues to the WNBA at its founding in 1997, the variables-of-interest in models one and two--calendar year and season number--might have differed in both result and significance. We interpret the null result of the third model as indicating no such divergence in interpretation, but rather that the two specifications of time are quite redundant.

Our main model finds that the same overall decrease in transitivity also holds within each league, with the exception of the WNBA. The fourth model is a post hoc regression in response to the WNBA's disparate trendline. Applying our primary model to only the WNBA data, we find that the positive trendline apparent in Figure 2 is not statistically different from 0.

The overall finding is consistent with our first hypothesis, that team-level advances in how to play basketball have created a divergence in performance among the ecosystem of teams and fewer competitive outcomes. This interpretation is supported by a corroborative analysis which shows an increase in the gap in winning percentage between the strongest and weakest teams, across all leagues ($t(100)=2.328$, $p<0.05$; Figure 3; Table 3). The regression, with the standard deviation of the winning percentage as the dependent variable and standard errors robust for heteroskedasticity, shows that the distribution of wins in an average season seems to be less equal over time.

[[Figure 2 here]]

[[Table 1 here]]

[[Figure 3 here]]

[[Table 2 here]]

When disaggregating by league, the NBA's longevity allows for the most interpretability. A random selection of three games relating three teams has resulted in a lower frequency of transitive, competitive outcomes relative to random baseline. The same trend shows for the more limited tenures of the BAA and the ABA. Competition in all three major men's professional basketball associations demonstrates a trend toward lower levels of parity, at least in the regular season. These results could be due to any of the many consequences of the professionalization of basketball over seventy years, including changes in practices, salaries, recruitment practices, and talent pools. To return to the analogy of regulatory capture, between some sub-group of "super-teams," a group of "tanking" teams, or a complicit league, certain teams seem to have sufficiently overcome the league's regulatory processes to institutionalize an advantage that manifests in a reduction of the unpredictability of the sport. The WNBA may or may not be an exception to this trend.

**V. Discussion**

Capture and other symptoms of runaway rich-get-richer dynamics are evident in every field: organization science, industrial economics, sociology and social network analysis, and political science. It is not bad by definition, but can, in each domain, be associated with the emergence of a qualitatively new type of overpowered agent. The main implication of our work is that these runaway dynamics are strong enough to emerge even in the presence of a single, high-powered administrator with a responsibility, and incentives, to prevent that emergence. We cannot conclude from this work how the administrator failed. Do teams develop strategies to insulate themselves from the administrator's authority? Do dominating teams bring enough resources to the league that it becomes willing to sacrifice parity for revenue? A better understanding of mechanisms for capture will lend insight into the difficult problem of preventing the runaway accumulation of resources endemic to so many social systems in so many disciplines.

While significant research inside and outside of sports has measured the interactions and relationships of competitive teams in a network, another line of investigation that is ready for attention is that of the orchestrating agents in the organizing system, whom we have called the market makers. The market maker exists as an interested party, with distinct goals and motivations from that of teams. As the teams in each professional basketball league vie for sustained dominance, leagues work in the opposite direction, attempting to ensure that contests remain unpredictable. The market makers, the league administrators, have a mandate and substantial power to protect the competitive balance of the network, promote ongoing innovation, and nurture the game's appeal to basketball's audience, the public. Many innovations in basketball mitigate the ability of dominant teams to achieve complete dominance, but the tendency toward the pooling of resources has permitted decreased competitiveness among teams, however modest, despite the efforts of the leagues.

### VI. Limitations and Future Work

While the overall trendlines, being linear fits, seem continuous, it is worth considering whether there are jumps in transitivity rate that correspond to league-level rule changes. A few of the more prominent rule-changes may have had an impact on competitiveness. The introduction of the shot-clock in 1954 markedly increased scoring. The three-point line was installed in 1977. The draft has undergone multiple reforms, most significantly in 1989 and most recently in 2019. Instant replay was introduced in 2002 and has since expanded. SportVU cameras were installed in 10 arenas in 2012, but then gained league-wide installation the following year. We do not consider these possibilities quantitatively, but, at least to the eye, any discontinuities associated with these dates do not stand out. Our proposition is that there is insufficient evidence of any single innovation disrupting the overall predictability trends. We do not find reason to reject that the effects of changes in basketball are largely incremental across time.

At the conceptual level, we have favored a framing for this work that emphasizes the relationship between a network of competitors and the entity in charge of setting the terms of the iterative contests. Other mechanisms exist that could have explained this result with other outcomes which could have been consistent with a successful system for producing and diffusing innovation.

For example, our behavioral and mechanistic focus masks broader social or institutional factors that may be driving our result. Sports teams exist to make money, by attracting viewers, by hosting compelling contests. League administrators do enforce top-down regulations to promote rebalancing, and while we have measured the effectiveness against the "ceiling" of competitive balance, this study does not measure league administrator effectiveness against a completely unregulated environment. Further analysis could measure the differential between the results found here and a hypothetical, completely unmonitored sporting competition, a competitive "floor."

Alternatively, since skill levels in sports face a potential ceiling effect at the upper limit of human performance, some parity due to the reduced skill differences near the limit might mask capture of the market maker by the teams: if teams are already operating near the performance limits of humans in the sport, then rampant institutionalized power may not be evident in observed differentials. Innovation may be diffusing, and skill levels may be so close as to create idiosyncratic outcomes, but a sport league can still trend toward capture if the league administrators lose regulatory power over how teams accumulate resources. In this scenario, we simply selected the wrong dependent measure.

Last, the method employed in this study falls in a distinct category between the in-game analysis of competitiveness using scoring events and the macro-scale analysis of team performance across decades (Ben-Naim et al., 2005; Price et al. 2010; Gabel and Redner, 2012; Merritt and Clauset, 2013, 2014; Martín-González et al., 2016; Rockerbie, 2016). If results of similar studies arrive at different findings — and this work has alternatively identified the NBA as being overall more or less competitive than other sports depending on the unit of analysis used — the general method for measuring the competitiveness of a contest in any domain invites comparisons and a better understanding of when and how managing institutions successfully implement the diffusion of innovation.

## VII. Conclusion

At every level of society, market makers structure competitive environments to harness uncertainty and unpredictability toward the generation of positive, network-wide benefits. At the same time, the teams they regulate work to gain an institutional advantage by gaining leverage over their market maker. This study utilizes data from professional basketball results to determine the degree to which team competitions, observed over decades, exhibit symptoms of "regulatory capture" by a subset of increasingly powerful teams, despite the active efforts of the manager class to maintain parity. Through the use of a transitivity violation test we determined that professional basketball exhibits decreased competitiveness amongst teams, suggesting some degree of success by individual competitors within the market to condition and capture regulator activity. By measuring the imbalanced competitive relationships among contestants, this study suggests that capture is a phenomenon that is ubiquitous enough to appear even in systems that are managed by a powerful agent who is motivated to prevent it.

## References

Abrahamson, E. Managerial Fads and Fashions: The Diffusion and Rejection of Innovations. *The Academy of Management Review* **16**, 586-612. doi:10.5465/amr.1991.4279484 (1991).

Ahn, Y. Y., Han, S., Kwak, H., Moon, S. & Jeong, H. Analysis of Topological Characteristics of Huge Online Social Networking Services. *Proceedings of the 16$^{th}$ international conference on World Wide Web*. doi:10.1145/1242572.1242685 (2007).

Barabási, A. L. & Albert, R. Emergence of Scaling in Random Networks. *Science* **286**, 509-512. doi:10.1126/science.286.5439.509 (1999).

Barnett, H. G. *Innovation: The Basis of Cultural Change.* New-York, Toronto, London (1953).

Barnett, W. P. & Hansen, M. T. The red queen in organizational evolution. *Strategic management journal* **17**, 139-157 (1996).

Barney, J. Firm Resources and Sustained Competitive Advantage. *Journal of Management* **17**, 99-120. doi:10.1177/014920639101700108 (1991).

Ben-Naim, E., Redner, S., & Vázquez, F. What is the most Competitive Sport? arXiv preprint physics/0512143 (2005).

Blind, K. The Influence of Regulations on Innovation: A Quantitative Assessment for OECD Countries. *Research Policy* **41**, 391-400. doi:10.1016/j.respol.2011.08.008 (2012).

Clauset, A., Arbesman. S., & Larremore, D. B. Systematic inequality and hierarchy in faculty hiring networks. *Sci Adv*. **1**, doi: 10.1126/sciadv. (2015): e1400005.

Dion, M. L., Sumner, J. L., & Mitchell, S. M. Gendered Citation Patterns Across Political Science and Social Science Methodology Fields. *Political Analysis* **26**, 312-327. doi:10.1017/pan.2018.12 (2018).

Encaoua, D., & Jacquemin, A. Degree of monopoly, indices of concentration and threat of entry. *International economic review* **21**, 87-105. Retrieved from http://www.econis.eu/PPNSET?PPN=390566667 (1980).

Gabel, A. & Redner, S. Random Walk Picture of Basketball Scoring. *Journal of Quantitative Analysis in Sports* **8**. doi:10.1515/1559-0410.1416 (1980).

Hausman, J. A. & Leonard, G. K. Superstars in the National Basketball Association. *Journal of Labor Economics* **15**, 586-624. http://www.econis.eu/PPNSET?PPN=261058320 (1997).

Hsieh, C.T & Klenow, P. Misallocation and Manufacturing TFP in China and India. NBER Working Paper **13290**, http://econpapers.repec.org/paper/nbrnberwo/13290.htm (2007).

Kogan, L., Papanikolaou, D., Seru, A., & Stoffman, N. Technological Innovation, Resource Allocation, and Growth. *The Quarterly Journal of Economics* **132**, 665-712 (2017).

Martín-González, J.M., de Saá Guerra, Y., García-Manso, J.M, Arriaza, E., & Valverde-Estévez, T. The Poisson Model Limits in NBA Basketball: Complexity in Team Sports. *Physica A: Statistical Mechanics and its Applications* **464**, doi:10.1016/j.physa.2016.07.028 (2016).

Merritt, S. & Clauset, A. Environmental Structure and Competitive Scoring Advantages in Team Competitions. *Scientific Reports* **3**, 3067. doi:10.1038/srep03067 (2013).

———Scoring Dynamics Across Professional Team Sports: Tempo, Balance and Predictability. *EPJ Data Science* 3: 4. doi:10.1186/epjds29 (2014).

Michels, R. *Political Parties: A Sociological Study of the Oligarchical Tendencies of Modern Democracy*. Hearst's International Library Company (1915).

Org. for Economic Cooperation & Development (OECD*). Oslo manual: Guidelines for collecting and interpreting innovation data*. No. 4. (2005).

Pareto, V. *The Mind and Society*. Harcourt, Brace & Company: New York (1935).

Price, J., Soebbing, B. P., Berri, D., &, Humphreys, B.R. Tournament Incentives, League Policy, and NBA Team Performance Revisited. *Journal of Sports Economics* **11**, 117-135. doi:10.1177/1527002510363103 (2010).

Radicchi, F., Fortunato, S., & Castellano, C. Universality of Citation Distributions: Toward an Objective Measure of Scientific Impact. *Proceedings of the National Academy of Sciences* **105**, 17268-17272. doi:10.1073/pnas.0806977105 (2008).

Robertson, T. S. The Process of Innovation and the Diffusion of Innovation. *Journal of Marketing* **31**, 14-19. doi:10.1177/002224296703100104 (1967).

Rockerbie, D. W. Exploring Interleague Parity in North America: The NBA Anomaly. *Journal of Sports Economics*, **17**, 286-301. (2016).

Simon, H. A., & Bonini, C. P. The size distribution of business firms. *The American economic review*: 607-617 (1958).

Stallman, Richard. The GNU Manifesto. *Dr. Dobb's Journal of Software Tools* (1985).

Stenseth, N. C., & Maynard Smith, J. Coevolution in Ecosystems: Red Queen Evolution Or Stasis? *Evolution* **38**, 870-880. doi:10.1111/j.1558-5646.1984.tb00358.x (1984).

Ugander, J., Karrer, B., Backstrom, L. & Marlow, C. 2011. The Anatomy of the Facebook Social Graph. arXiv preprint arXiv:1111.4503.

Westney, A. SportVU tracking system could find reach beyond court for NBA, teams. *Sports Business Daily*, October 29 (2013).

Wojnarowski, A. & Lowe, Z.Why teams may be apprehensive about enforcing tampering. *ESPN.com*, September 19 (2019).

**Acknowledgements**

The authors wish to thank George Barnett and Scott Traum for their contributions to this work.

**Author's Contributions**

A.S. and S.F. contributed to the design and implementation of the research, the analysis of the results and the writing of the manuscript.

**Statement of Interests**

The authors declare no competing interests.

Tables and Figures

*Table 1: A regression shows separate season- and year-sensitive effects on transitivity over time. The dependent variable is the transitivity violation rate: the difference between observed and baseline intransitivity). Model 1 measures the difference in transitivity violation between the observed and baseline. For each season, transitivity violations drop at a statistically significant level. The same occurs when the calendar year is substituted for the season number as seen in Model 2. Using both time variables eliminates the effect of either, suggesting redundancy. Given the disparate trendline for the WNBA, the authors conducted an isolated post-hoc regressions, seen in model 4, but found that the seasonal change in the rate of intransitivity does not significantly differ from 0 either when measuring based on the number of years since founding, or when accounting for the calendar year.*

|  | **Season only** | **Year only** | **Season*Year** | **WNBA only** |
|---|---|---|---|---|
| **Season since league founding** | -0.0003* (0.0001) |  | -0.0218 (0.0169) | -0.0004 (0.0012) |
| **Year** |  | -0.0003* (0.0002) | -0.0004 (0.0002) |  |
| **Season*Year** |  |  | 0.0000 (0.0000) |  |
| **ABA** | -0.0068 (0.0101) | -0.0008 (0.0088) | -0.0062 (0.0103) |  |
| **BAA** | -0.0208* (0.0081) | -0.0218* (0.0085) | -0.0293* (0.0131) |  |
| **WNBA** | -0.0113 (0.0086) | 0.0045 (0.0085) | NA |  |
| **Intercept** | -0.0421*** (0.0069) | 0.5992 (0.3066) | 0.7511 (0.4755) | -0.0527** (0.0176) |
| **Degrees of Freedom** | 95 | 95 | 94 | 20 |
| **Adjusted $R^2$** | 0.011 | 0.011 | 0.016 | -0.044 |
| *p<0.05 **p<0.01 ***p<0.001 |  |  |  |  |

*Table 2: Regression results. The dependent variable is the standard deviation of winning percentage among teams in a league. Seasonal win-percentage distributions seem to have widened over time, substantiating the results of the transitivity violation test.*

|  | All leagues |
|---|---|
| **Season** | 0.0004* |
|  | (0.0002) |
| **ABA** | 0.0120 |
|  | (0.0118) |
| **BAA** | 0.0289** |
|  | (0.0086) |
| **WNBA** | 0.0234* |
|  | (0.0093) |
| **Intercept** | 0.1353*** |
|  | (0.0089) |
| **Degrees of Freedom** | 100 |
| **Adjusted R²** | 0.0604 |
| *$p<0.05$ | |
| **$p<0.01$ | |
| ***$p<0.001$ | |

*Figure 1: Transitivity implies a coherent ordering of teams based on performance: if A beats B and B beats C then A should beat C. The more this ordering is observed in triplets of games, the more imbalanced that competitive relationship is. By contrast, if all three teams defeat each other, the result is an intransitive relationship with no clear dominance. We would expect high transitivity in a more skill-based game such as chess, and low transitivity in a game of chance like roulette. Changes in a game's transitivity over time suggest a systematic change in that game's predictability. We observe how this measure changes over time as one game, professional basketball, evolves over seven decades, interpreting changes in transitivity in terms of the competing interests of teams and the leagues that coordinate them.*

## Transitivity

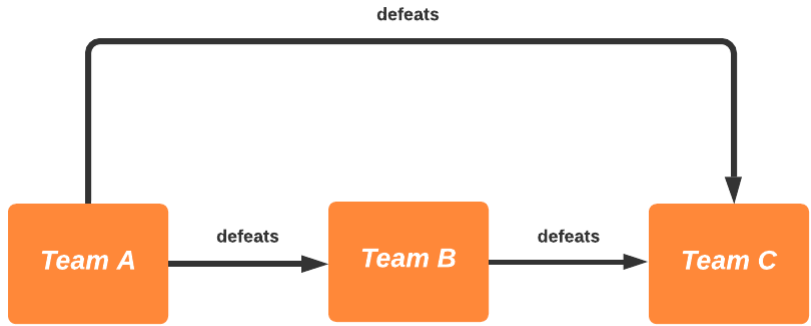

Team is dominant

Increase in *transitivity* suggests less parity and greater role of skill through increased competitive imbalance

## Transitivity Violation

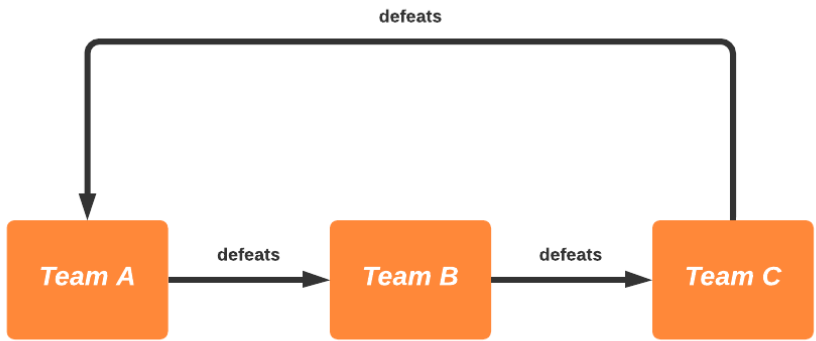

No team is dominant

Increase in *transitivity violations* suggests greater parity and competitiveness

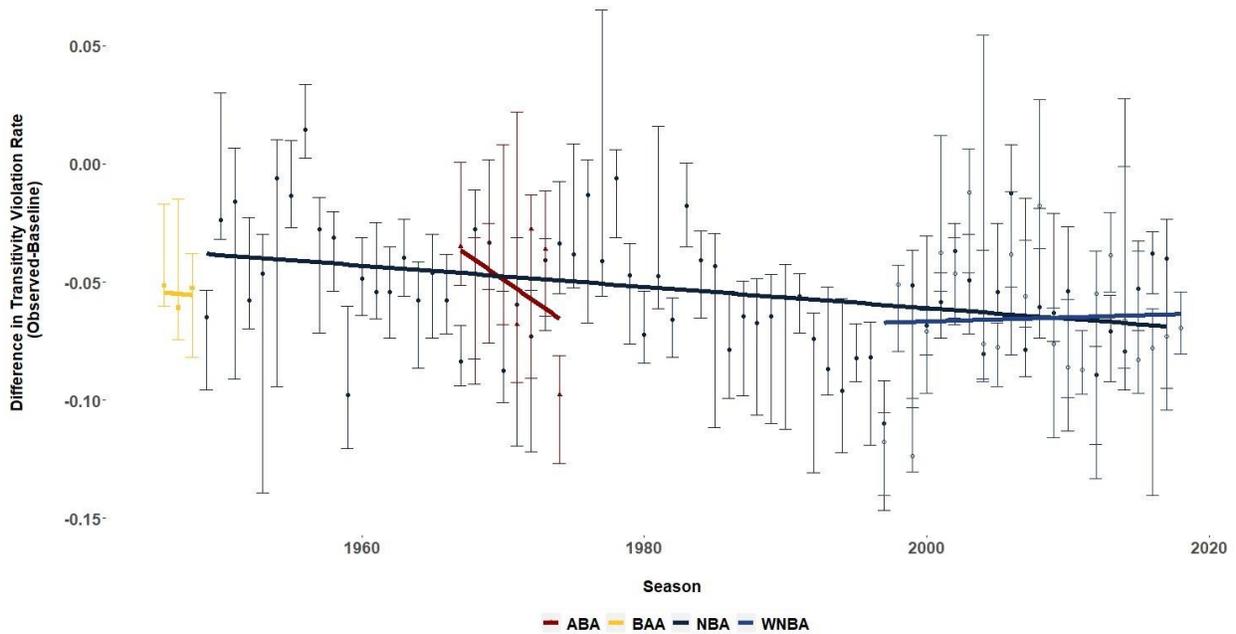

*Figure 2: Professional basketball has become less competitive over decades. This decrease in parity of the game is consistent with a failure of leagues to prevent some teams from hoarding competitive advantage. The y-axis measures the difference between the rate of transitivity violations in the observed set of basketball results minus the baseline rate of shuffled, permuted scores. With the exception of the WNBA, all professional basketball leagues feature a downward trendline, indicating transitivity violations are increasing as compared to a random baseline. With less intransitivity, basketball games feature less uncertainty, consistent with a divergence in the relative abilities of teams in each league, which is in turn consistent with a rich-get-richer dynamic that benefits top teams despite the supposed equalizing influence of the league.*

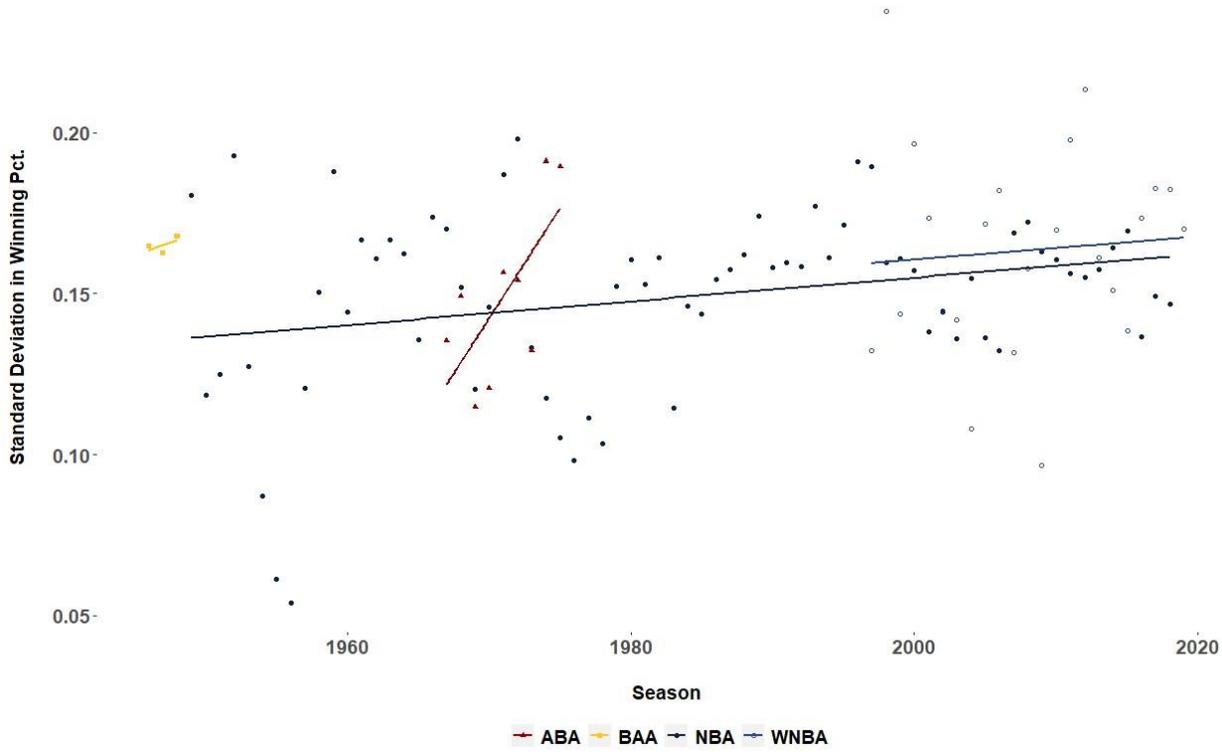

*Figure 3: Professional basketball features a wider distribution of competitive success in 2019 compared to 1946, as measured by the standard deviation of teams' seasonal winning percentage. This supports our argument that the increased determinism of basketball is due to the ability of top-performing teams to evade league control and institutionalize their competitive advantage.*

# Supplementary Information

*Supplementary figure 1. After shuffling the scores within each season to create a within-season baseline for intransitivity, we can compare the results of this randomized result dataset to the real observed competitive rate of transitivity violations. This figure shows the underlying data trend that is analyzed in Figure 2: Figure 2 plots the difference between the observed and baseline data shown here. The result suggests a gradual decrease in transitivity violations, suggesting a trend toward less parity, that is counterbalanced in part, but not completely, by an increase over time in the baseline violation rate.*

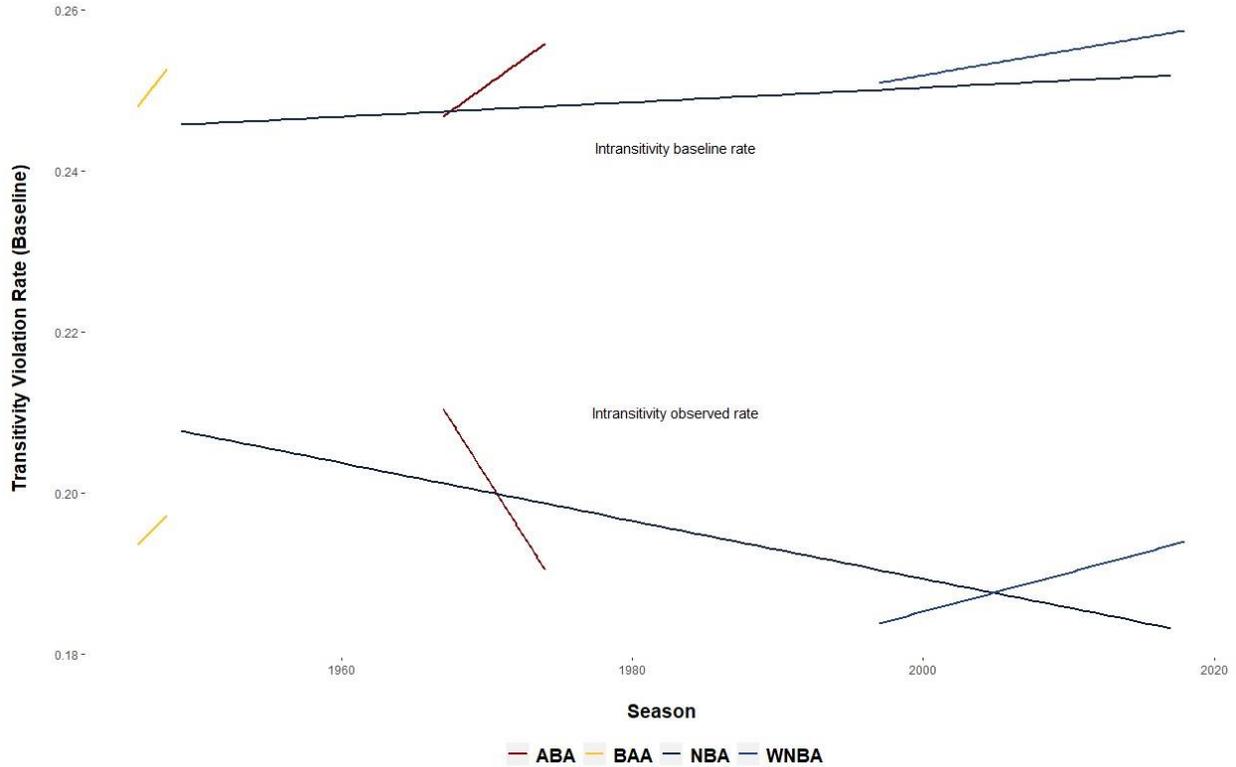